# Coding for Memory with Stuck-at Defects

Yongjune Kim and B. V. K. Vijaya Kumar
Electrical and Computer Engineering, Data Storage Systems Center (DSSC)
Carnegie Mellon University
Pittsburgh, USA
yongjunekim@cmu.edu, kumar@ece.cmu.edu

*Abstract*—In this paper, we propose an encoding scheme for partitioned linear block codes (PLBC) which mask the stuck-at defects in memories. In addition, we derive an upper bound and the estimate of the probability that masking fails. Numerical results show that PLBC can efficiently mask the defects with the proposed encoding scheme. Also, we show that our upper bound is very tight by using numerical results.

## I. INTRODUCTION

Most memory systems (e.g., flash memory, phase-change memory, etc.) exhibit two types of imperfections that threaten the data reliability. The first type is a defective memory cell, i.e., defect, whose cell value is stuck-at a particular value independent of the input. For example, some of the cells of a binary memory may be stuck-at 0, and when a 1 is attempted to be written into a stuck-at 0 cell, an error results. The second type of imperfection is a noisy cell which can occasionally result in a random error. The distinction between these two types of imperfections is that stuck-at defects are permanent, whereas errors caused by noise are intermittent. Often the terms hard and soft errors are used to describe stuck-at errors and noise-induced errors, respectively [1]–[4].

By carefully testing the memory, it is possible to know information about defects such as locations and stuck-at values, and this information can be exploited in the encoder and decoder for more efficient coding schemes. This problem was first addressed by Kuznetsov and Tsybakov [1]. They assume that the location and value of the defects are available to the encoder, but not to the decoder [1], [2]. Note that a typical scheme which uses the defects information in the decoder is the erasure decoding.

Later, Heegard proposed the partitioned linear block codes (PLBC) that efficiently incorporate the defect information in the encoding process and are capable of correcting both stuck-at errors (due to defects) and random errors [4]. Recently, his work has drawn attention for nonvolatile memories because flash memories and phase change memories (PCM) suffer from stuck-at defects [5]–[8].

In [4], the encoding algorithm of PLBC is stated as an implicit optimization problem to mask defects, i.e., find a codeword whose values at the locations of defects match the stuck-at values at those locations. If the number of defects $u$ is less than $d_0$ which is the minimum distance of PLBC, this optimization problem can be solved by Gaussian elimination (GE) and the masking always succeeds, i.e., $\Pr(\text{masking failure}) = 0$. Also, a polynomial interpolation encoding scheme with reduced computational complexity has been proposed for partitioned cyclic codes (PCC) [5].

However, the PLBC encoding algorithm for more than $d_0 - 1$ defects is an optimization problem with exponential computational complexity [8]. Although an encoding scheme based on cross entropy method has been recently proposed [8], this cross entropy encoding scheme cannot guarantee the success of masking defects even if $u \leq d_0 - 1$. In addition, it is difficult to obtain analytical results such as upper bound because the cross entropy encoding scheme uses random samples.

In this paper, we propose a simple two-step scheme for encoding PLBC. The proposed two-step encoding scheme can efficiently mask the defects and has much better performance than the conventional PLBC without the need to solve an optimization problem. In addition, it guarantees the success of masking defects for $u \leq d_0 - 1$.

We will also derive a tight upper bound on the probability of masking failure by using the weight distribution of the code. In addition, the estimate of $\Pr(\text{masking failure})$ will also be derived for $d_0 \leq u \leq d_0 + \lfloor (d_0 - 1)/2 \rfloor$ (where $\lfloor x \rfloor$ is the largest integer not greater than $x$). The derived analytical expressions for the upper bound and the estimate are very important for data storage systems such as nonvolatile memories. The reason is that data storage systems have very high requirement in reliability, e.g., BER $\leq 10^{-15}$ and it is very difficult to obtain simulation results for such very low BERs and we have to rely on analytical results. These analytical results are the main contribution of our paper.

The rest of the paper is as follows. Section II explains the channel model and PLBC. In section III, the optimization problem in the PLBC encoding is explained and two-step encoding scheme is proposed. In section IV, the upper bound on the probability of masking failure is derived and the numerical results of section V show the performance of two-step encoding scheme and the tightness of the upper bound. Section VI concludes the paper.

## II. CODING FOR MEMORY WITH STUCK-AT DEFECTS

### A. Channel Model for Memory with Stuck-at Defects

In [3], [4], the channel model for memories with defects has been introduced. The model assumes both stuck-at defects and additive random errors. We will use the channel model of [4].

Let $q$ be a power of a prime and $F_q$ be the Galois field with $q$ elements. Let $F_q^n$ denote the set of all $n$-tuples over $F_q$. Define an additional variable "$\lambda$" that denotes "non-defect" state, $\tilde{F}_q = F_q \cup \lambda$ and define the "$\circ$" operator $\circ: F_q \times \tilde{F}_q \to F_q$ by

$$x \circ s = \begin{cases} x, & \text{if } s = \lambda; \\ s, & \text{if } s \neq \lambda. \end{cases} \quad (1)$$

An $n$-cell memory with defects and random errors is modeled by

$$y = (x \circ s) + z \quad (2)$$

where $x$ is the vector to be stored, $z$ is the random error vector and $s$ is the defect vector. The addition "+" is defined over the field $F_q$ and both + and $\circ$ operate on the vectors componentwise.

The number of defects $u$ is equal to the number of non-$\lambda$ components in $s$, and the number of random errors is defined by $t = \|z\|$ where $\|\cdot\|$ is the Hamming weight of the vector.

A stochastic model for the generation of defects and random errors in memory cells is obtained by assigning probabilities to the defect and random error events. The $(\varepsilon, p)$ $q$-symmetric discrete memoryless memory cell ($q$-SDMMC) is modeled by the equation $Y = (X \circ S) + Z$, where $X, Y, Z \in F_q$, $S \in \tilde{F}_q$,

$$\Pr(S = s) = \begin{cases} 1 - \varepsilon, & s = \lambda; \\ \dfrac{\varepsilon}{q}, & s \neq \lambda, \end{cases}$$
$$\Pr(Z = z | S = \lambda) = \begin{cases} 1 - p, & z = 0; \\ \dfrac{p}{q - 1}, & z \neq 0. \end{cases} \quad (3)$$

Fig. 1 illustrates the channel model for memories with defects when $q = 2$. We will focus on the channel model of Fig. 1.

*B. Partitioned Linear Block Codes*

In [4], Heegard proposed the $[n, k, l]$ PLBC which is a pair of linear subspaces $\mathcal{C}_1 \subset F_q^n$ and $\mathcal{C}_0 \subset F_q^n$ of dimension $k$ and $l$ such that $\mathcal{C}_1 \cap \mathcal{C}_0 = \{0\}$. Then the direct sum

$$\mathcal{C} \triangleq \mathcal{C}_1 + \mathcal{C}_0 = \{c = c_1 + c_0 | c_1 \in \mathcal{C}_1, c_0 \in \mathcal{C}_0\} \quad (4)$$

is an $[n, k + l]$ linear block code (LBC) with a generator matrix $G = [G_1^T \; G_0^T]^T$ where $G_1$ generates $\mathcal{C}_1$ and $G_0$ generates $\mathcal{C}_0$ (superscript $T$ denotes transpose). The parity check matrix $H$ is $r \times n$ matrix with $k + l + r = n$. A message inverse matrix $\tilde{G}_1$ is defined as $k \times n$ matrix with $G_1 \tilde{G}_1^T = I_k$ (the $k$-dimensional identity matrix), and $G_0 \tilde{G}_1^T = 0_{l,k}$ (the $l \times k$ zero matrix) [4].

The encoding and decoding algorithms of PLBC are as follows [4].

*Encoding:* To encode a message $w \in F_q^k$ into a codeword $c = w G_1 + d G_0$ where $d \in F_q^l$ is chosen to minimize $\|(c \circ s) - c\|$.

*Decoding:* Retrieve $y = (c \circ s) + z$. Compute the syndrome $v = y H^T$ and choose $\hat{z} \in F_q^n$ which minimizes $\|z\|$ subject to $z H^T = v$. Then $\hat{w} = \hat{c} \tilde{G}_1^T$ where $\hat{c} = y - \hat{z}$.

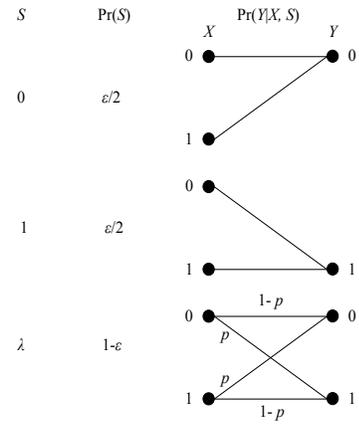

Fig. 1. Channel model for binary memories with defects [3].

A pair of minimum distances $(d_1, d_0)$ of an $[n, k, l]$ PLBC are given by

$$\begin{aligned} d_1 &= \min_{\substack{c \tilde{G}_1^T \neq 0 \\ c H^T = 0}} \|c\|, \\ d_0 &= \min_{\substack{c \neq 0 \\ c G_0^T = 0}} \|c\| \end{aligned} \quad (5)$$

where $d_1$ is greater than or equal to the minimum distance of the $[n, k + l]$ LBC with parity check matrix $H$, while $d_0$ is the minimum distance of the $[n, k + r]$ LBC with parity check matrix $G_0$ [4]. For convenience, we will define $d_0$ as the minimum distance of PLBC.

*Theorem 1 [4]:* An $[n, k, l]$ PLBC with minimum distances $(d_1, d_0)$ is a $u$-defect, $t$-error correcting code if

$$u < d_0 \text{ and } 2t < d_1$$

or

$$u \geq d_0 \text{ and } 2(u + t + 1 - d_0) < d_1.$$

If $u < d_0$, all defects will be successfully masked and $\|(c \circ s) - c\| = 0$. Otherwise, it may be that $\|(c \circ s) - c\| \neq 0$ which results in masking failure. The unmasked defects will be regarded as random errors in the decoder.

*C. Partitioned Cyclic Codes*

An $[n, k, l]$ partitioned cyclic code (PCC) is a more restrictive class of PLBC. An $[n, k, l]$ PCC has two generator polynomials, $g_1(x)$ of degree $r$ ($n = k + l + r$) and $g_0(x)$ of degree $k + r$ such that $g_1(x) | g_0(x)$ and $g_0(x) | x^n - 1$. The encoding and decoding algorithms of PCC are as follows [4].

*Encoding:* A codeword of PCC is $c(x) = w(x) g_1(x) + d(x) g_0(x)$ where $w(x)$ is a message and $d(x)$ is chosen to minimize $\|(c(x) \circ s(x)) - c(x)\|$.

*Decoding:* Receive $y(x) = (c(x) \circ s(x)) + z(x)$ and compute the syndrome $v(x) = y(x) \bmod g_1(x)$. Choose $\hat{z}(x) \in F_q^n(x)$ which minimizes $\|z(x)\|$ subject to $z(x) \bmod g_1(x) = v(x)$. Then

$$\hat{w}(x) = \frac{(y(x) - \hat{z}(x)) \bmod g_0(x)}{g_1(x)}.$$

The partitioned Bose, Chaudhuri, and Hocquenghem (PBCH) code is a special class of PCC. The generator polynomials and minimum distances can be designed by a similar method as standard BCH codes [4].

### III. Encoding Scheme for Partitioned Linear Block Codes

#### A. Optimization Problem of Encoding Algorithm

The encoding of PLBC includes an implicit optimization problem which can be formulated as follows [4], [5], [8].

$$\begin{aligned} \boldsymbol{d}^* &= \arg\min_{\boldsymbol{d}} \|\boldsymbol{d} G_0^{\Psi_u} + \boldsymbol{w} G_1^{\Psi_u} - \boldsymbol{s}^{\Psi_u}\| \\ &= \arg\min_{\boldsymbol{d}} \|\boldsymbol{d} G_0^{\Psi_u} + \boldsymbol{b}^{\Psi_u}\| \end{aligned} \quad (6)$$

where $\boldsymbol{b}^{\Psi_u} = \boldsymbol{w} G_1^{\Psi_u} - \boldsymbol{s}^{\Psi_u}$ and $\Psi_u = \{i_1, \cdots, i_u\}$ indicates the locations of $u$ defects. Also, we define $\boldsymbol{s}^{\Psi_u} = [s_{i_1}, \cdots, s_{i_u}]$, $G_1^{\Psi_u} = [\boldsymbol{g}_{1,i_1}, \cdots, \boldsymbol{g}_{1,i_u}]$, and $G_0^{\Psi_u} = [\boldsymbol{g}_{0,i_1}, \cdots, \boldsymbol{g}_{0,i_u}]$ where $\boldsymbol{g}_{1,i}$ and $\boldsymbol{g}_{0,i}$ are the $i$-th columns of $G_1$ and $G_0$ respectively. The $G_0^{\Psi_u}$ is $l \times u$ matrix and $\boldsymbol{b}^{\Psi_u}$ is $u$-tuple row vector.

If $u < d_0$, $\text{rank}(G_0^{\Psi_u})$ is always $u$ because of (5). Thus, the solution $\boldsymbol{d}^*$ satisfying $\boldsymbol{d} G_0^{\Psi_u} = \boldsymbol{b}^{\Psi_u}$ (corresponding to masking success) can be always obtained via Gaussian elimination [4] or some other solution method for linear equations. In addition, for PCC, a polynomial interpolation encoding scheme can be applied for reduced computational complexity [5].

However, if $u \geq d_0$, the optimal solution $\boldsymbol{d}^*$ may fail to mask all defects and the computational complexity for solving the optimization problem is exponential, which is impractical as $l$ increases [8]. In [4], a modified formulation of (6) was described, which chooses only $\min(d_0 - 1, u)$ locations among $u$ defects instead of solving the complex optimization problem. Then, the solution of the modified formulation can always be solved. We name it as one-step encoding scheme for comparison with our two-step encoding scheme. The one-step encoding scheme is summarized as follows.

| One-step encoding scheme |
|---|
| • Obtain $m = \min(d_0 - 1, u)$ |
| • Choose $m$ locations among $u$ defects: $\Psi_m = \{i_1, \cdots, i_m\}$. |
| • Use the equation $\boldsymbol{d} G_0^{\Psi_m} = \boldsymbol{b}^{\Psi_m}$ and find the solution $\boldsymbol{d}$. |

For PBCH codes, we found that it is desirable to choose $(d_0 - 1)$ locations in descending order from higher degree element. The reason is that the unmasked defects of higher degree than $g_0(x)$ result in error multiplication during the decoding operation of $(y(x) - \hat{z}(x)) \bmod g_0(x)$.

#### B. Two-Step Encoding Scheme

For the standard form, $\boldsymbol{d} G_0^{\Psi_u} = \boldsymbol{b}^{\Psi_u}$ can be expressed by

$$(G_0^{\Psi_u})^T \boldsymbol{d}^T = (\boldsymbol{b}^{\Psi_u})^T \quad (7)$$

where $(G_0^{\Psi_u})^T$ is $u \times l$ matrix, $\boldsymbol{d}^T$ is $l$-tuple column vector and $(\boldsymbol{b}^{\Psi_u})^T$ is $u$-tuple column vector. (7) has at least one solution if and only if

$$\text{rank}\left((G_0^{\Psi_u})^T\right) = \text{rank}\left((G_0^{\Psi_u})^T | (\boldsymbol{b}^{\Psi_u})^T\right) \quad (8)$$

where $\left((G_0^{\Psi_u})^T | (\boldsymbol{b}^{\Psi_u})^T\right)$ is the augmented matrix [9].

If $u < d_0$, (7) always has at least one solution because $d_0$ is the minimum distance of the LBC with parity check matrix $G_0$, which means that any $d_0 - 1$ columns of $G_0$ are linearly independent, i.e., $\text{rank}(G_0^{\Psi_u}) = u$. This condition of $u < d_0$ corresponds to Theorem 1.

However, even if $u \geq d_0$, it is possible that $\text{rank}(G_0^{\Psi_u}) = u$ and (7) can have at least one solution. In addition, even if $\text{rank}(G_0^{\Psi_u}) < u$, (7) can have at least one solution so long as (8) holds, which will be explained in Lemma 2.

The two-step encoding scheme will be as follows.

| Two-step encoding scheme |
|---|
| Step 1: |
| • Try to solve (7). |
|   - If $u < d_0$, the solution $\boldsymbol{d}$ to (7) will always exist and go to end. |
|   - If $u \geq d_0$, the solution $\boldsymbol{d}$ to (7) can be obtained so long as (8) holds. |
|     ▪ If we can obtain the solution, go to end. |
|     ▪ Otherwise, go to step 2. |
| Step 2: |
| • Choose $(d_0 - 1)$ locations among $u$ defects: $\Psi_{d_0-1} = \{i_1, \cdots, i_{d_0-1}\}$. |
| • Use the equation $\boldsymbol{d} G_0^{\Psi_{d_0-1}} = \boldsymbol{b}^{\Psi_{d_0-1}}$ and find the solution $\boldsymbol{d}$. |
| End |

If (7) can be solved in Step 1, the number of unmasked defects will be zero, i.e., $\|(\boldsymbol{c} \circ \boldsymbol{s}) - \boldsymbol{c}\| = 0$. If $\boldsymbol{d}$ is obtained in Step 2, $0 < \|(\boldsymbol{c} \circ \boldsymbol{s}) - \boldsymbol{c}\| \leq u - (d_0 - 1)$. For PBCH codes, we will choose $(d_0 - 1)$ locations in descending order from higher degree element in step 2 for the same reason as in one-step encoding scheme.

If the computational complexity of one-step encoding scheme with GE is $O(\min(d_0 - 1, u)^3)$, the complexity of two-step encoding scheme with GE will be $O(u^3)$. Thus, the computational complexity of the two-step encoding scheme is much less than exponential computational complexity for solving the optimization problem when $u \geq d_0$.

### IV. Upper Bound of Partitioned Linear Block Codes

Assuming that the two-step encoding scheme has been used, we will derive the upper bound for $\Pr(\text{masking failure}) = \Pr(\|(\boldsymbol{c} \circ \boldsymbol{s}) - \boldsymbol{c}\| \neq 0)$. In addition, the exact estimate of $\Pr(\text{masking failure})$ will be obtained for $u \leq d_0 + \lfloor (d_0 - 1)/2 \rfloor$.

*Lemma 2:* For random data, $\Pr(\text{masking failure})$ is given by

$$\Pr(\text{masking failure}) = \sum_{j=1}^{u-1} \frac{2^j - 1}{2^j} \Pr(\text{rank}(G_0^{\Psi_u}) = u - j). \quad (9)$$

*Proof:* The augmented matrix $\left(\left(G_0^{\Psi_u}\right)^T | (b^{\Psi_u})^T\right)$ of (8) can be transformed into the reduced row echelon form. If $\text{rank}(G_0^{\Psi_u}) = u$, the linear equation of (7) will be always solved and $\Pr(\text{masking failure}) = 0$. If $\text{rank}(G_0^{\Psi_u}) = u - j$ for $j > 0$, the last $j$ rows of the reduced row echelon form of $\left(G_0^{\Psi_u}\right)^T$ are zero vectors. In order to satisfy the condition of (8), the last $j$ elements of the column vector $(b^{\Psi_u})^T$ should be zeros and $\Pr(\text{masking failure})$ will be $(2^j - 1)/2^j$ for random data. Thus, $\Pr(\text{masking failure})$ will be as in (9). □

From Lemma 2, the lower bound and the upper bound on $\Pr(\text{masking failure})$ are given by

$$\frac{1}{2} \cdot \Pr(\text{rank}(G_0^{\Psi_u}) < u) \leq \Pr(\text{masking failure}) \leq \Pr(\text{rank}(G_0^{\Psi_u}) < u). \quad (10)$$

*Lemma 3:* The upper bound on $\Pr(\text{rank}(G_0^{\Psi_u}) < u)$ is given by

$$\Pr(\text{rank}(G_0^{\Psi_u}) < u) \leq \frac{\sum_{w=1}^{u} A_w \binom{n-w}{u-w}}{\binom{n}{u}} \quad (11)$$

where $A_w$ is the number of codewords of Hamming weight $w$ in the LBC with parity check matrix $G_0$. Note that this LBC is the dual code of $\mathcal{C}_0$.

*Proof:* We will define $\Psi_w(c)$ as the locations of nonzero elemtents of $c$ as follows.

$$\Psi_w(c) = \{i | c_i \neq 0\} \quad (12)$$

where $w = \|c\|$ and $c_i$ is the $i$-th element of $c$. For example $\Psi_4(c) = \{1, 3, 4, 5\}$ for $c = (1011100)$.

If there exists a nonzero codeword $c$ such that $\Psi_w(c) \subseteq \Psi_u$ where $c$ is a codeword in LBC with parity check matrix $G_0$, then $\text{rank}(G_0^{\Psi_u}) < u$. The reason is that $G_0^{\Psi_{w(c)}}$ which is a submatrix of $G_0^{\Psi_u}$ and the columns of $G_0^{\Psi_{w(c)}}$ are not linearly independent since $G_0 c^T = 0$.

If $w > u$, it is impossible that $\Psi_w(c) \subseteq \Psi_u$. We do not need to consider $c$ such that $\|c\| > u$.

If $w = u$, the only possible condition of $\Psi_w(c) \subseteq \Psi_u$ is $\Psi_w(c) = \Psi_u$. Therefore, the number of possible $\Psi_u$ such that $\text{rank}(G_0^{\Psi_u}) < u$ will be $A_u$.

If $w < u$, it is possible that $\Psi_w(c) \subset \Psi_u$. For any $c$ such that $\Psi_w(c) \subset \Psi_u$, the number of possible $\Psi_u$ will be $\binom{n-w}{u-w}$ since $w$ locations are fixed. Thus, for $w < u$, the number of all possible $\Psi_u$ such that $\text{rank}(G_0^{\Psi_u}) < u$ is less than or equal to $\sum_{w=1}^{u-1} A_w \binom{n-w}{u-w}$ because of double counting. Double counting occurs when $\Psi_u$ includes the locations of nonzero elements of at least two codewords.

The number of all possible $\Psi_u$ will be $\binom{n}{u}$. Therefore, the upper bound on $\Pr(\text{rank}(G_0^{\Psi_u}) < u)$ will be (11). □

*Theorem 4:* The upper bound on $\Pr(\text{masking failure})$ is given by

$$\Pr(\text{masking failure}) \leq \frac{\sum_{w=1}^{u} A_w \binom{n-w}{u-w}}{\binom{n}{u}} \quad (13)$$

which agrees with Theorem 1 because $\sum_{w=1}^{u} A_w \binom{n-w}{u-w} = 0$ for $u < d_0$.

*Proof:* By (10) and Lemma 3, the upper bound on $\Pr(\text{masking failure})$ is given by (13). □

*Lemma 5:* For $u \leq d_0 + \lfloor (d_0 - 1)/2 \rfloor = d_0 + t_0$, $\Pr(\text{masking failure})$ is given by

$$\Pr(\text{masking failure}) = \frac{1}{2} \cdot \frac{\sum_{w=1}^{u} A_w \binom{n-w}{u-w}}{\binom{n}{u}} \quad (14)$$

where $t_0 = \lfloor (d_0 - 1)/2 \rfloor$ is the error correcting capability of the LBC with parity check matrix $G_0$.

*Proof:* The proof has two parts. First, we will show that $\Pr(\text{rank}(G_0^{\Psi_u}) < u) = \sum_{w=1}^{u} A_w \binom{n-w}{u-w}/\binom{n}{u}$ for $u \leq d_0 + t_0$, which means that there is no double counting in Lemma 3. Second, we will prove that $\Pr(\text{rank}(G_0^{\Psi_u}) < u) = \Pr(\text{rank}(G_0^{\Psi_u}) = u - 1)$ for $u \leq d_0 + t_0$, which means that $\Pr(\text{rank}(G_0^{\Psi_u}) \leq u - 2) = 0$. Taking into account (9) and these two parts, $\Pr(\text{masking failure})$ will be given by

$$\Pr(\text{masking failure}) = \frac{1}{2} \cdot \Pr(\text{rank}(G_0^{\Psi_u}) = u - 1)$$
$$= \frac{1}{2} \cdot \Pr(\text{rank}(G_0^{\Psi_u}) < u)$$
$$= \frac{1}{2} \cdot \frac{\sum_{w=1}^{u} A_w \binom{n-w}{u-w}}{\binom{n}{u}}.$$

In order to show that $\Pr(\text{rank}(G_0^{\Psi_u}) < u) = \sum_{w=1}^{u} A_w \binom{n-w}{u-w}/\binom{n}{u}$ for $u \leq d_0 + t_0$, let us consider any two codewords $c_1$ of weight $w_1$ and $c_2$ of weight $w_2$ ($d_0 \leq w_1 \leq w_2$) in LBC with parity check matrix $G_0$. From the definition of (12), the locations of nonzero elements of these two codewords are given by

$$\Psi_{w_1}(c_1) = \{i_{1,1}, \cdots, i_{1,w_1}\}$$

and

$$\Psi_{w_2}(c_2) = \{i_{2,1}, \cdots, i_{2,w_2}\}.$$

Let $\Psi_\alpha = \{i_1, \cdots, i_\alpha\}$ denote $\Psi_{w_1}(c_1) \cap \Psi_{w_2}(c_2)$. Since the order of elements in $\Psi_{w_1}(c_1)$ and $\Psi_{w_2}(c_2)$ is not important, without loss of generality, we can claim that

$$\Psi_{w_1}(c_1) = \{\Psi_\alpha, i_{1,1}, \cdots, i_{1,\beta_1}\}$$

and

$$\Psi_{w_2}(c_2) = \{\Psi_\alpha, i_{2,1}, \cdots, i_{2,\beta_2}\}$$

where $i_{1,j}$ (for $j \in \{1, \cdots, \beta_1\}$) and $i_{2,j}$ (for $j \in \{1, \cdots, \beta_2\}$) are re-indexed locations considering $\Psi_\alpha$. Note that $\beta_1 = w_1 - \alpha$ and $\beta_2 = w_2 - \alpha$.

If we set $c_3 = c_1 + c_2$, then $\|c_3\| = w_3 = \beta_1 + \beta_2$. Since $c_3$ is also a codeword due to the property of linear codes, the following conditions should be true because of the definition of $d_0$.

$$\alpha + \beta_1 \geq d_0, \alpha + \beta_2 \geq d_0, \text{ and } \beta_1 + \beta_2 \geq d_0$$

Therefore, $2(\alpha + \beta_1 + \beta_2) \geq 3d_0$ and $\alpha + \beta_1 + \beta_2 \geq d_0 + \lfloor (d_0 + 1)/2 \rfloor$ because $\alpha + \beta_1 + \beta_2$ is integer.

For double counting to occur, there exist two codewords $\boldsymbol{c}_1$ and $\boldsymbol{c}_2$ such that $\Psi_{w_1}(\boldsymbol{c}_1) \cup \Psi_{w_2}(\boldsymbol{c}_2) \subseteq \Psi_u$. It means that double counting can occur only when $u \geq \alpha + \beta_1 + \beta_2 \geq d_0 + \lfloor (d_0 + 1)/2 \rfloor$. Therefore, there is no double counting when $u \leq d_0 + \lfloor (d_0 - 1)/2 \rfloor = d_0 + t_0$.

Now, we will show that $\Pr(\text{rank}(G_0^{\Psi_u}) < u) = \Pr(\text{rank}(G_0^{\Psi_u}) = u - 1)$ for $u \leq d_0 + t_0$. If $\text{rank}(G_0^{\Psi_u}) < u$, $\Psi_{w_1}(\boldsymbol{c}_1) \subseteq \Psi_u$ for at least one codeword $\boldsymbol{c}_1$ of weight $w_1 \geq d_0$. Without loss of generality, we can assume that $\Psi_u = \{i_1, \cdots, i_u\} = \{i_1, \cdots, i_{u-w_1}, \Psi_{w_1}(\boldsymbol{c}_1)\}$ where $\Psi_u \backslash \Psi_{w_1}(\boldsymbol{c}_1) = \{i \in \Psi_u | i \notin \Psi_{w_1}(\boldsymbol{c}_1)\} = \{i_1, \cdots, i_{u-w_1}\}$.

In order to satisfy $\Pr(\text{rank}(G_0^{\Psi_u}) < u) = \Pr(\text{rank}(G_0^{\Psi_u}) = u - 1)$, the following two conditions should be hold.

(i) $\Psi_{w_j}(\boldsymbol{c}_j) \not\subseteq \Psi_{w_1}(\boldsymbol{c}_1)$ for any codeword $\boldsymbol{c}_j \neq \boldsymbol{c}_1$.

If $\Psi_{w_j}(\boldsymbol{c}_j) \subset \Psi_{w_1}(\boldsymbol{c}_1)$ for $\boldsymbol{c}_j$ of weight $w_j < w_1$, $\text{rank}(G_0^{\Psi_{w_1}(\boldsymbol{c}_1)})$ can be less than $w_1 - 1$, which makes $\text{rank}(G_0^{\Psi_u}) < u - 1$. If $\Psi_{w_1}(\boldsymbol{c}_1) = \Psi_{w_j}(\boldsymbol{c}_j)$, then $\boldsymbol{c}_1 = \boldsymbol{c}_j$.

(ii) Any column among $G_0^{\Psi_u \backslash \Psi_{w_1}(\boldsymbol{c}_1)} = [\boldsymbol{g}_{0,i_1}, \cdots, \boldsymbol{g}_{0,i_{u-w_1}}]$ should be linearly independent of other columns of $G_0^{\Psi_u}$.

If any column among $G_0^{\Psi_u \backslash \Psi_{w_1}(\boldsymbol{c}_1)}$ is a linear combination of other columns of $G_0^{\Psi_u}$, then $\text{rank}(G_0^{\Psi_u}) < u - 1$ although the previous condition (i) holds. The reason is that both $G_0^{\Psi_{w_1}(\boldsymbol{c}_1)}$ and $G_0^{\Psi_u \backslash \Psi_{w_1}(\boldsymbol{c}_1)}$ have at least one dependent column. In this case, $\Psi_u$ will include the locations of nonzero elements of at least two codewords.

From (i) and (ii), we can see that $\text{rank}(G_0^{\Psi_u}) < u - 1$ if $\Psi_u$ includes the locations of nonzero elements of at least two codewords. For $\text{rank}(G_0^{\Psi_u}) = u - 1$, $\Psi_u$ should include the locations of nonzero elements of only one codeword, which is same as the condition of no double counting. We have already shown that there is no double counting for $u \leq d_0 + t_0$. Thus, $\text{rank}(G_0^{\Psi_u}) = u - 1$ for $u \leq d_0 + t_0$. □

*Theorem 6:* $\Pr(\text{masking failure})$ is given by

$$\Pr(\text{masking failure}) = \begin{cases} 0, & u < d_0; \\ \frac{1}{2} \cdot \frac{\sum_{w=1}^{u} A_w \binom{n-w}{u-w}}{\binom{n}{u}}, & d_0 \leq u \leq d_0 + t_0; \\ \leq \frac{\sum_{w=1}^{u} A_w \binom{n-w}{u-w}}{\binom{n}{u}}, & u > d_0 + t_0. \end{cases} \quad (15)$$

*Proof:* Combining Theorems 1, 4, and Lemma 5, $\Pr(\text{masking failure})$ is given by (15). □

It is worth mentioning that the difference between (13) and (14) is not significant because $\Pr(\text{masking failure})$ is generally considered in log scale.

Until now, it was assumed that the number of defects $u$ is fixed. We can assume that the number of defects in a codeword follows the binomial distribution because $q$-SDMMC of (3) is independent and identically distributed (i.i.d.). Then, the upper bound on Pr(masking failure) is given by

$$\begin{aligned}
&\Pr(\text{masking failure}) \\
&\leq \sum_{u=d_0}^{n} \binom{n}{u} \varepsilon^u (1-\varepsilon)^{n-u} \cdot \frac{\sum_{w=1}^{u} A_w \binom{n-w}{u-w}}{\binom{n}{u}} \\
&= \sum_{u=d_0}^{n} \varepsilon^u (1-\varepsilon)^{n-u} \cdot \sum_{w=1}^{u} A_w \binom{n-w}{u-w}.
\end{aligned} \quad (16)$$

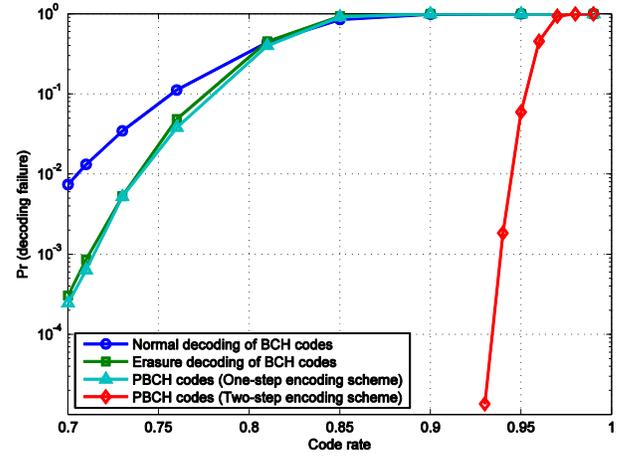

Fig. 2. Comparison of Pr(decoding failure) for nomal decoding of BCH codes, erasure decoding of BCH codes, and one-step encoding scheme of PBCH codes, and two-step encoding scheme of PBCH codes ($n = 1023, \varepsilon = 40/1023 \approx 3.91 \times 10^{-2}, p = 0$).

## V. NUMERICAL RESULTS

We will compare the normal decoding of BCH codes, the erasure decoding of BCH codes, and the PBCH coding schemes. The normal decoding of BCH codes means that there is no information about the defects, and the erasure decoding uses the locations of defects in the decoder. The PBCH coding schemes use the defects information in the encoder.

Fig. 2 shows $\Pr(\text{decoding failure})$ of the normal decoding and the erasure decoding of conventional BCH codes as well as $\Pr(\text{masking failure})$ of the PBCH coding schemes. In regard to PBCH codes, we used $[n, k, l]$ PBCH codes with $r = 0$ which means $d_1 = 0$. Therefore, $\Pr(\text{masking failure})$ is equivalent to $\Pr(\text{decoding failure})$. We assumed that $n = 1023, \varepsilon = 40/1023 \approx 3.91 \times 10^{-2}$, and $p = 0$.

$\Pr(\text{decoding failure})$ of the normal decoding without any information of defects is the worst. $\Pr(\text{decoding failure})$ of the erasure decoding and that of the one-step encoding scheme of PBCH codes are almost same because the condition for masking success of one-step encoding scheme in Theorem 1 is same as the condition for decoding success of the erasure decoding, i.e., $u < d_0$. $\Pr(\text{decoding failure})$ of the two-step encoding scheme is significantly better, showing that (7) can be solved in many cases even if $u \geq d_0$.

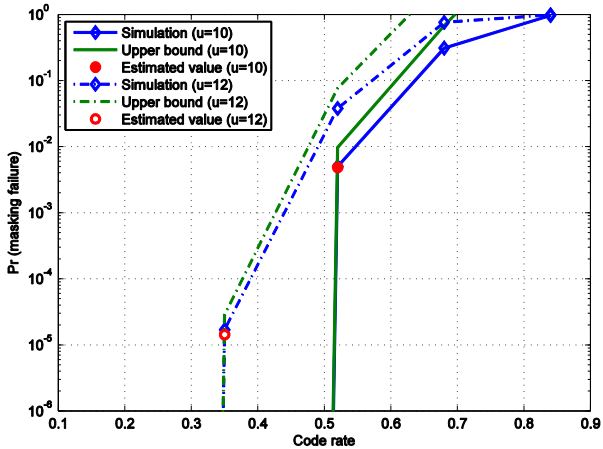

Fig. 4. Comparison of simulation results, upper bound, and estimates ($n = 31, p = 0$).

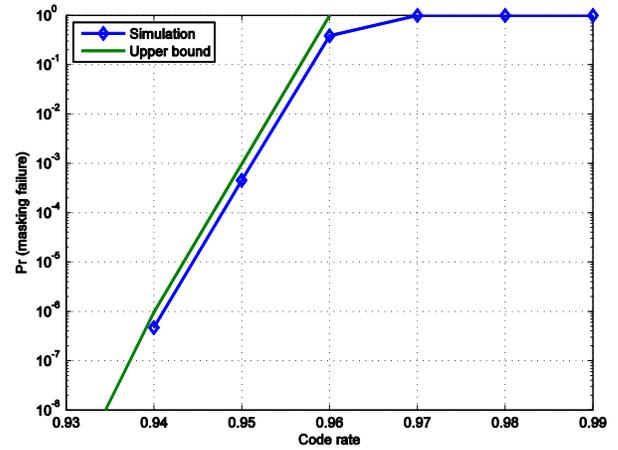

Fig. 3. Comparison of simulation results and upper bound ($n = 1023, u = 40, p = 0$). Estimates are not displayed because they are lower than $10^{-20}$.

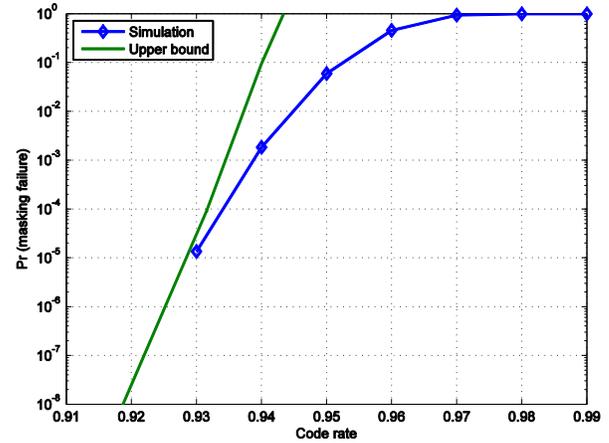

Fig. 5. Comparison of simulation results and upper bound ($n = 1023, \varepsilon = 40/1023 \approx 3.91 \times 10^{-2}, p = 0$).

Fig. 4 shows that the upper bound of (13) is close to the simulation results for $\Pr(\text{masking failure})$. In addition, the estimated values of (14) are well matched with simulation results of $\Pr(\text{masking failure})$. We see that our upper bound approaches $\Pr(\text{masking failure})$ and is eventually same as $\Pr(\text{masking failure})$ as the code rate decreases (i.e., $d_0$ increases). We used ($n = 31$) PBCH codes and $A_w$ was calculated by MacWilliams identity [10].

Fig. 3 shows the result of ($n = 1023$) PBCH codes and $u = 40$. When $n$ is large, computing $A_w$ is intractable. Thus, we used the approximation of $A_w \approx \frac{1}{2^{n-k}} \binom{n}{w}$ for $w \geq d_0$ [10]. In spite of this approximation, our upper bound is very close to the simulation results. The estimates for $d_0 \leq u \leq d_0 + \lfloor (d_0 - 1)/2 \rfloor$ are not displayed because they are lower than $10^{-20}$. Fig. 5 shows that the upper bound of (16) is also close to the simulation results.

## VI. CONCLUSION

In this paper, we showed that PLBC can have very good performance with a simple two-step encoding scheme instead of solving the optimization problem with exponential computational complexity. We derived the upper bound of $\Pr(\text{masking failure})$ for $u > d_0 + t_0$ and the estimate of $\Pr(\text{masking failure})$ for $d_0 \leq u \leq d_0 + t_0$.